\documentclass{ar-1col}

\usepackage[left=3cm,right=3cm,top=2cm,bottom=2cm]{geometry}

\usepackage{url}
\usepackage{comment}
\usepackage{orcidlink}
\usepackage[numbers]{natbib}

\jname{Annu. Rev. Condens. Matter Phys.}
\jvol{AA}
\jyear{YYYY}
\doi{10.1146/((please add article doi))}

\begin{document}

\markboth{Le Verge-Serandour $\cdot$ Alim}{\textit{Physarum polycephalum}: Smart network adaptation}

\title{\textit{Physarum polycephalum}: Smart network adaptation}

\author{Mathieu Le Verge-Serandour\orcidlink{0000-0002-3819-5099}, Karen Alim\orcidlink{0000-0002-2527-5831}
\affil{Technical University of Munich, Germany; TUM School of Natural Sciences, Department of Bioscience; Center for Protein Assemblies (CPA); email: k.alim@tum.de}
}

\begin{abstract}
Life evolved organisms to adapt dynamically to their environment and autonomously exhibit behaviours. While complex behaviours in organisms are typically associated with the capability of neurons to process information, the unicellular organism \textit{Physarum polycephalum} disabuses us by solving complex tasks despite being just a single although gigantic cell shaped into a mesmerizing tubular network. In \textit{Physarum}, smart behaviours arise as network tubes grow or shrink due to the mechanochemical coupling of contractile tubes, fluid flows and transport across the network. Here, from a physicist's perspective, we introduce the biology and active chemo-mechanics of this living matter network. We then review \textit{Physarum}'s global response in migration and dynamic state to its environment before revisiting its network architecture and flow and transport patterns. Finally, we summarize recent studies on storing and processing information to mount well-informed behaviours.
\end{abstract}

\maketitle


\section{INTRODUCTION}
\begin{figure}[h!]
    \centering
    \includegraphics[scale=0.4]{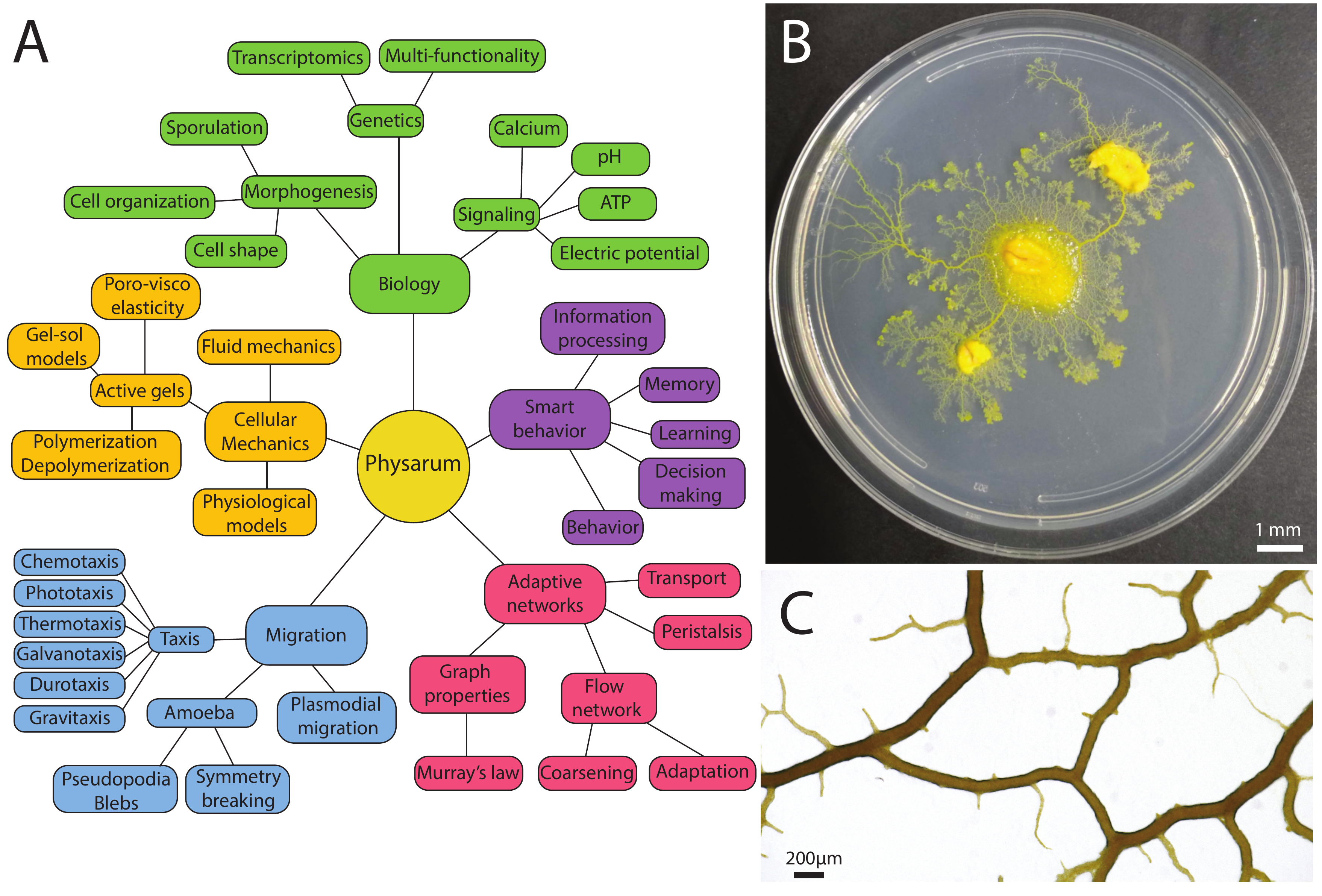}
    \caption{\textbf{\textit{Physarum} at different scales}. (A) Graphical abstract of the sections and fields discussed in the review. (B) \textit{Physarum polycephalum} plated on 1.5\% Agar in a petri dish, fed with oat flakes. (C) Closer picture of \textit{Physarum}'s tube with color camera.
   }
    \label{fig:introduction}
\end{figure}

\textit{Physarum polycephalum}, a giant single-cell slime mould, fascinates researchers with its sophisticated behaviour despite its simple built. The network-shaped body plan of \textit{Physarum} plasmodia tops typical cell size reaching up to meters in length while enclosing thousands even millions of nuclei \citep{Kauffman1975}, which allows for complex behaviour similar to multi-cellular organisms. A life form that beat the odds of 600 million years of evolution \citep{Kjellin.2021} to still thrive on earth today and combine traits of what later on became animals, plants and fungi in itself \citep{Schaap2016}. First described 200 years ago in 1822 \citep{Schweinitz1834}, biologists studied \textit{Physarum} as model for differentiation and development, motility and cell cycle \citep{Sauer1982_book, Rusch1966, Raub1982, Alim.2013cb}. At the last turn of the century the discovery of \textit{Physarum}'s ability to adapt its network morphology, see Fig. \ref{fig:introduction}B, to solves complex task like finding the shortest path through a maze \citep{Nakagaki:2000kp} or conquering the Steiner tree problem \citep{Tero:2010bx} excited computer scientists and physicists alike. These observations suggested that \textit{Physarum} is somehow a ''smart'' organism, amenable to quantitative studies as the inherently two-dimensional semi-transparent network tubes, see Fig. \ref{fig:introduction}C, allow live imaging of network dynamics and the fluid flows pervading the network and shuffling organelles and biochemical reagents around. Physical parameters such as fluid flow velocity are easy to vary non-invasively as flow velocities scale with organism size that reaches from hundreds of micrometers to typically tens of centimeters. Moreover a rich repertoire of adversive or attractive stimuli is already documented, allowing to probe the physics of behaviour emerging from the mechanochemical interactions of living matter.

In this review, we introduce the biological background of \textit{Physarum} and its mechanical and biochemical make-up from a physicist perspective before we summarize \textit{Physarum}'s dynamics regarding migration, network reorganization and behaviour, all summarised in the graphical abstract of Fig. \ref{fig:introduction}A.

\begingroup
\let\clearpage\relax
\section{BIOLOGICAL BACKGROUND}

\begin{figure}[h!]
    \centering
    \includegraphics[scale=0.4]{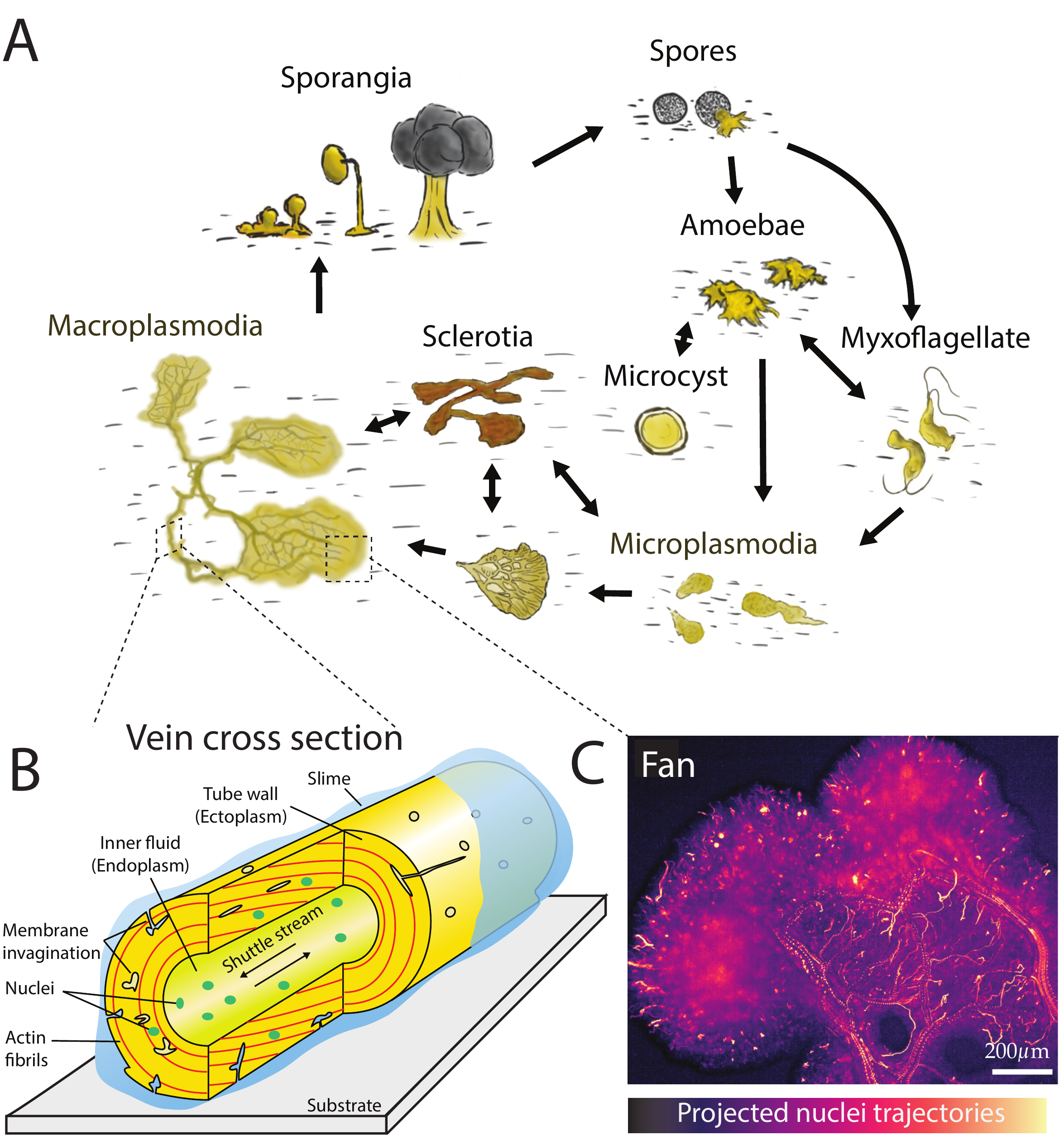}
    \caption{\textbf{\textit{Physarum} developmental cycle and cellular organization}. (A) Developmental cycle starts with haploid spores that germinate and evolve into amoeba or flagellates. Upon fusion of two cells, the then diploid cell becomes a plasmodia, growing with food into its striking network shape. Desiccation or lack of nutrients induce differentiation into sclerotia or sporangium that produces spores. The network-shaped plasmodia has two distinct regions: (B) veins, composing most of the network, made of an invaginated tube wall and inner fluid transporting nuclei with a shuttle stream ; (C) growing fronts, or so-called fans, expand with cytoplasmic flow displacing organelles and projection of nuclei trajectories (SYTO 62-labelled), adapted from \citep{Gerber2022}.
   }
    \label{fig:biology}
\end{figure}

    \subsection{Physarum developmental cycle}
\textit{Physarum polycephalum} belongs to the Myxogastria, a class of slime molds of the Mycetozoa phylum, in the kingdom of protists \citep{Fiore-Donno2010}. 
\textit{Physarum} shares a common ancestor with Amoebozoa and Metazoa \citep{Schaap2016, Glockner2008}. The many morphologies of \textit{Physarum} throughout its developmental cycle \citep{Marwan2001} is one of its fascinating features, see Fig.~\ref{fig:biology}A. Starting from mature spores, a haploid \textbf{amoebae} hatches. This amoeba can crawl, feed on bacteria, and divide as single-celled organisms \citep{Sauer1982_book}. Depending on external conditions, an amoeba transforms either into a biflagellated \textbf{myxoflagellate}, or a \textbf{microcyst} - a walled cell, obtained under unfavorable conditions such as high salt concentration \citep{Hara1981} - or two amoebae may fuse and give a diploid \textbf{zygote} \citep{Sauer1982_book, Stockem1994}. The zygote will continue developing, with nuclear divisions and organism growth but without cell division (cytokinesis), rendering it the so-called \textbf{plasmodial} stage. This multinucleated unicellular stage is also referred to as a syncytium. At the plasmodial stage, the organism continues developing and growing as nutrients are provided, from 100 µm diameter \textbf{micro-plasmodia} to the millimeter-sized mature \textbf{macro-plasmodia} \citep{Oettmeier2018}. 
Upon exposure to light (UV or visible), a starving plasmodium becomes a \textbf{sporangium} \citep{Glockner2017} its mass cleaves and differentiates into fruiting bodies containing hundreds of haploid spores, spreading at the outburst of the bodies. Spores encountering favorable conditions will form haploid amoeba re-starting its developmental cycle.
Otherwise, with desiccation or lack of nutrients \citep{Patterson1977}, the plasmodial individual enters a new phase: the \textbf{sclerotia}, a dormant stage with thick cellulose walls \citep{Ogawa2010}. It can remain in the sclerotia stage for several months, even years, until rehydration and the addition of nutrients revive it back into the plasmodial stage \citep{Sauer1982_book, Oettmeier2017}. 

\subsection{Cellular organization}
The most studied stages of development of \textit{Physarum} are the amoeba and micro- and macro-plasmodium stages. Despite the very different forms of these stages, all share common cellular structures, which we briefly review here. The \textbf{protoplasm} refers to the cell's living part surrounded by a plasma membrane. It encompasses two physical phases which can be converted into one another \citep{Isenberg1976}: the \textbf{endoplasm} is the inner part of the protoplasm and is surrounded by the \textbf{ectoplasm}, the outer part, see Fig.~\ref{fig:biology}B. Both are made of cytoskeletal proteins, but they differ in their mechanical properties. The ectoplasm, also termed ``tube wall'', is a poroelastic structure and generates most of the contractile forces of the cell, while the endoplasm, or ``inner fluid'', has low viscosity, and its flow allows the dispersion of nutrients, organelles, and chemicals throughout the organism \citep{Oettmeier2018}.
The macro-plasmodia is a vascular network of well-defined connected tubes. The tips are often referred to as \textbf{fans} and are protrusions of a thin protoplasmic sheet, which inflates and deflates with the contractions of the upstream network and as the cytoplasmic flow is pushed into them, see Fig.~\ref{fig:biology}C. Some fans are retracted as \textit{Physarum} migrates on surface, while others persist and eventually form channels to become tubes as the rest of the network.

Surrounding the whole organism is the \textbf{slime}, a layer of mucopolysaccharides, secreted through exocytosis \citep{Sesaki1997}. This outer layer consists of a large portion of the plasmodium volume and is remarkably left as a ``trace'' of the past presence of the migrating plasmodia \citep{Reid2012, Reid2013, Huynh2017}. It is composed of several carbohydrates (mainly glucose), proteins, and sulfate groups \citep{Huynh2017, McCormick1970}. In particular, the slime coating contains antimicrobial and antifungal compounds, acting as a protective layer from the environment \citep{McCormick1970, Wolf1981_II, Patino-Ramirez2019}.

The plasma membrane, or \textbf{plasmalemma}, of \textit{Physarum} is the outer part of the cell. The plasma membrane exhibits many invaginations \citep{Wohlfarth-Bottermann1974}, connected to the surface through pores with an average diameter of 1.7 µm and an average distance of 7.3 µm apart \citep{Oettmeier2018}. The invaginations considerably increase the surface area of \textit{Physarum}, which raises the food uptake of the organism. The number of invaginations depends on the nutrition of the substrate: the higher the food uptake, the higher the volume of invaginations of the plasmalemma on nutritional veins, as opposed to transporting veins \citep{Achenbach1979}. The plasma membrane regenerates rapidly, in 5-6 seconds \citep{Wohlfarth-Bottermann1970}.

Beneath the plasma membrane is the actin cytoskeleton, which consists roughly 20\% of the total plasmodial mass, with 60\% of it being filamentous F-actin \citep{Stockem1994}, located in the tube wall. At the same time, the endoplasm contains a lot of G-actin monomers, and the polymerization of G- to F-actin controls the endoplasm-to-ectoplasm transformation \citep{Isenberg1976}.
Besides, several actin-binding proteins have been reported: fragmin \citep{Hasegawa1980}, profilin \citep{Ozaki1984}, and most importantly, myosin, forming the actomyosin complex. The \textbf{actomyosin cortex} is one of the prominent components of \textit{Physarum}'s cytoskeleton and is responsible for the contractility and integrity of cell shape, and is found in two forms: a cortical and a fibrillar system \citep{Brix1987_VII}.
The cortical system is a dense meshwork located beneath the plasma membrane, in the tube wall, with varying thickness, from 0.5-0.7 µm to several micrometers at fans \citep{Stockem1994}. It ensures the stability of the cell surface and its invaginations.
The actomyosin fibrils are long sarcomeric bundles that propagate tension and contractility of the organism \citep{Ishigami1987, Wohlfarth-Bottermann1976}, and are responsible for the membrane invaginations \citep{Kukulies1984, Gawlitta1980}. The fibrils are helicoidally and longitudinally wrapping the tubes, with an average of 21µm in length \citep{Stockem1994, Oettmeier2018}. They also ensure adhesion to the substrate through pseudopodia \citep{Brix1987_VII}. Disruption of the actomyosin cortex with latrunculin A leads to the dilation of the cell and its disruption because of the high hydrostatic pressure of the cytoplasm \citep{Oettmeier2018}. Finally, microtubules have also been observed in \textit{Physarum} in the cytoskeleton at the plasmodial stage, with a parallel orientation to the long axis of the veins \citep{Salles-Passador1991}. Their role in both transport of vesicles and organelles, or their contribution to the cytoskeleton mechanics, remains poorly characterized. 
            
Cell organelles move freely within the cytoplasm, transported by the shuttle stream \citep{Oettmeier2018}. In \textit{Physarum}, one can find mitochondria, ribosomes, a poorly developed Golgi apparatus, and endoplasmic reticulum. It also contains many vacuoles filled with slime, food, or water \citep{Sauer1982_book}. The yellow color of \textit{Physarum} comes from several pigments: physarochrome, polycephalins, and chrysophysarin \citep{Steffan1987, Eisenbarth2000}, stored in pigment granules \citep{Sauer1982_book}. However, white mutant strains were also used to study the phototactic response of the organism \citep{Schreckenbach1981}.
            
    \subsection{Nuclei \& genetics}
Akin to fungal hyphae, \textit{Physarum} is a syncytium: a single cell with one nucleus at the amoeba stage, and millions of nuclei at the plasmodial stage \citep{Kauffman1975, Mela2020}. The diploid nucleus has a spherical shape of 3-6µm \citep{Sauer1982_book, Wohlfarth-Bottermann1983, Howard1932, Goodman2012}. Ultrastructural microscopy reveals that \textit{Physarum} shares several characteristics of eukaryotic cells: the nucleus is made of a nuclear envelope composed of lamina proteins \citep{Bekers1981}, with nuclear pores \citep{Sauer1982_book} and fibrils that may be actomyosin fibers or tubulin \citep{Lachapelle1987}, the latter being useful for the intranuclear mitosis \citep{Tanaka1973}.

In the amoeba stage, nuclear division, is an open mitosis with astral microtubules like in animals or plants. However, at the plasmodial stage, division consists of intranuclear mitosis, as in fungi \citep{Tanaka1973}. At this stage, nuclei are reported to  divide synchronously every 10 hours within 5 minutes, even for large plasmodia, up to 5 cm long \citep{Howard1932,Guttes1961,Kubbies1983}. Even more striking is the synchronization of the mitotic cycle after the fusion of two distinct plasmodia, with the new division period being dependent on the relative size of fused individuals \citep{Rusch1966}.

The shuttle stream of the cytoplasm displaces organelles and nuclei throughout the plasmodium \citep{Wohlfarth-Bottermann1983}. The distribution of nuclei was measured, with a high density of nuclei at the growing fronts compared to the veins. Two dynamic states of nuclei are found: mobile nuclei, transported by the shuttle stream, and immobile nuclei, trapped in the tube wall, that can, however, be released back into the inner fluid \citep{Gerber2022}, see Fig.~\ref{fig:biology}C. This trapping mechanism was proposed recently to be used by the unicellular organism to achieve local control of gene expression to adapt its response to the environment \citep{Gerber2022}.
However, the coupling of advection and the complex architecture of the plasmodia vascular network allows efficient mixing of nutrients and molecules \citep{Haupt2020}, but also of nuclei, which adds another layer of complexity: distant, isolated parts of the organism may communicate through the dispersion, and transported nuclei may ``seed'' newly growing fronts \citep{Gerber2022}.

\textit{Physarum}'s genome was recently sequenced: about 31,000 genes have been identified \citep{Schaap2016, Glockner2008}. In particular, a study of the transcriptome shows that two individuals from the same culture but exposed to different environments express different genes. Furthermore, within an individual, two regions, such as growing fronts and network vein, can express different genes, favoring, for example, actin polymerization or calcium-binding proteins \citep{Gerber2022}.
The mechanisms leading to nuclei trapping or release, long-range genetic communication, and local gene expression represent a fascinating research direction to study how multi-functionality can be achieved in a single cell.

Overall, \textit{Physarum} is a unique biological system providing many opportunities to study how multifunctionality may have appeared in unicellulars, with complex morphogenesis and able to reproduce many features of living species. For physicists, the ease of culture and ability to entirely image it represents a considerable advantage for modelling over other biological systems.

\section{MECHANICS}
In this section, we develop the concepts used to characterize the mechanical properties of \textit{Physarum}. First, we study the actomyosin cortex and its viscoelastic contractile properties, at the core of the cytoplasmic fluid pumping. Then, we discuss calcium dynamics, the key to solving the contractile oscillations. Finally, we detail the fluid mechanics involved in the cytoplasmic flow within \textit{Physarum}'s tubular architecture.

\begin{figure}
    \centering
    \includegraphics[scale=0.4]{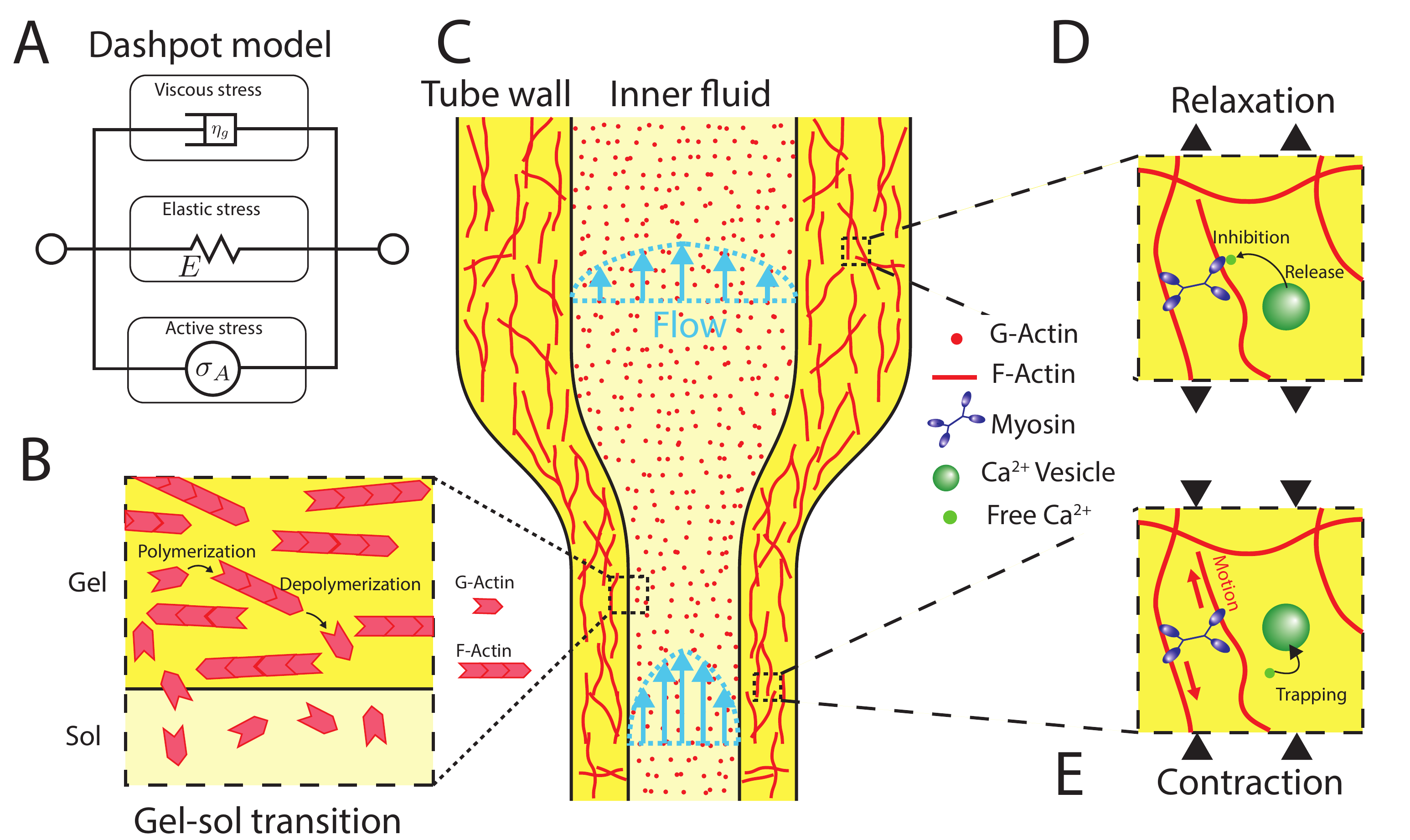}
    \caption{\textbf{Mechanics in \textit{Physarum}.} A vein (C) is made of a tube wall with F-actin filaments (gel) and flowing inner fluid (sol) containing G-actin monomers. The tube wall can be described (A) mechanically by a Kelvin-Voigt dashpot model with active stress, or (B) with a gel-sol model considering polymerization and depolymerization of G- to F-actin. The tube wall can (D) relax with the inhibition of myosin by calcium release from vesicles, or (E) contract by calcium trapping and activation of actomyosin. As a result, the tube diameter reduces, increasing locally the flow velocity.}
    \label{fig:mechanics}
\end{figure}
    
    \subsection{Mechanical models of \textit{Physarum}'s cytoskeleton}
    The cytoplasm can be divided into two phases: a viscous fluid phase, the \textbf{sol}, corresponding to the endoplasm or cytosol, and a poro-visco-elastic phase, the \textbf{gel}, corresponding to the ectoplasm of the tube walls \citep{Mogilner2018, Burla2019}, see Fig.~\ref{fig:mechanics}C.
In the case of \textit{Physarum}, the gel phase corresponds to a mesh of entangled proteins, filamentous actin (F-actin), interacting with myosin molecular motors as an active material to generate contractions \citep{Banerjee2020}, see Fig.~\ref{fig:mechanics}A.
The mechanical description of such material was proposed as an active gel theory \citep{Kruse2005, Joanny2009}, and applied to \textit{Physarum} \citep{Radszuweit2013}. For simplicity, we introduce the fundamental equations in one dimension. A section of the plasmodium is split into as sol and gel, with respective volume fractions $\phi_s$ and $\phi_g$, such that $\phi_s + \phi_g = 1$.
Without inertial forces, the force balance on a small volume element is written as $\partial_x \sigma+f = 0$, with $f$ the external forces and $\sigma$ the total stress. The total stress corresponds to the sum of the stress of the sol $\sigma_s$ and the stress of the gel $\sigma_g$, such that $\sigma = \phi_s \sigma_s + \phi_g \sigma_g$. The sol stress is often considered purely viscous $\sigma_s = \eta_s \partial_x v$, with $\eta_s$ the sol viscosity and $v$ the sol velocity field.
Constitutive laws are required to describe the gel stress and were first proposed for isotropic active visco-elastic material \citep{Oster1984}. Let $u$ be the displacement field of the gel. Considering the gel to be elastic at long time scales, the gel stress is the sum of viscous stress with gel viscosity $\eta_g$, elastic stress with $E$  Young's modulus and $\partial_x u$ the strain and active stress $\sigma_a$:  $\sigma_g = \eta_g \partial_x \dot{u} + E \partial_x u + \sigma_a$, which corresponds to the Kelvin-Voigt dashpot model shown in Fig.~\ref{fig:mechanics}A, \citep{Oster1984}. The active stress generates an active tension and depends on the local concentration of a chemical activator, likely calcium, discussed in section 3.2 \citep{Tero2005, Smith1994}. Other models of visco-elasticity have been debated and proposed to describe complex materials \citep{Teplov1991, Romanovskii1995, Alonso2017}. Young's modulus is measured around $E= 16.5$kPa, and gel viscosity about $\eta_g=7.5 \rm{kPa}.\rm{s}$ \citep{Fessel2018}, while the sol viscosity is around $\eta_s = 2$ Pa.s \citep{Sato1983}.
Including the hydrostatic pressure $p$ but neglecting osmotic pressure in the gel, one can finally write the force balance for the sol and the gel as
$0 = \partial_x \left(\eta_s \partial_x v - p\right) + f_s$, 
$0 = \partial_x \left(\eta_g \partial_x \dot{u} + E \partial_x u + \sigma_a - p\right) + f_g$, 
$0 = \partial_x \left( \phi_s v + \phi_g \dot{u} \right)$,
where the last equation corresponds to the incompressibility, i.e.,~total mass conservation. The boundary conditions are often no flux, Dirichlet conditions. The coupling of a chemical oscillator to the mechanics, through the active stress $\sigma_a$ of the cytoskeleton, produces mechanical oscillations \citep{Oster1984, Teplov1991}.

Such models have been refined, considering the cytoskeleton as a poro-visco-elastic gel interpenetrated by the viscous cytosol \citep{Mogilner2018, Dembo1986}. The flow of sol within the gel introduces frictional forces, which are derived from Darcy's law of porous media: $v = -\frac{\kappa}{\eta_s}\partial_x p$, where $\kappa$ is the permeability of the gel, in $\rm{m}^2$. The drag forces for the gel and the sol balance each other, such that $f_s = -f_g = \phi_g \phi_s \frac{\eta_s}{\kappa} (\dot{u}-v)$. Such formulations extended by mechano-chemical coupling, along with reaction-diffusion equations for the chemical oscillator, including an advection term generated by the local wall contractions \citep{Radszuweit2011, Radszuweit2014}. It was shown that the chemical does not necessarily need to be an autonomous oscillator to produce mechanical contractions \citep{Radszuweit2013}.
Additional forces can also account for the substrate adhesion required for locomotion \citep{Lewis2015}, and were included as friction forces of the gel on the substrate $f_{\rm{adh}} = \phi_g \gamma \dot{u}$, with $\gamma$ the substrate-gel drag coefficient \citep{Kulawiak2018, Kulawiak2019}. 
Just as active fluids \citep{Bois2011}, poroelastic models produce patterns and oscillations of the protoplasmic droplet's height, with spiraling, traveling, standing, and radial contractile waves reported \citep{Radszuweit2014}.
These models assume the sol-gel volume fractions to be constant, with no polymerization or depolymerization of actin proteins. Including the dynamics of polymerization/depolymerization as a net rate $J$ leads to the mass conservation equations
$\partial_t \phi_g + \partial_x(\phi_g \dot{u}) = J$
for the gel and 
$\partial_t \phi_s + \partial_x (\phi_s v) = -J$
for the sol. The ratio of G-actin to F-actin was shown to determine the viscosity of the cytoplasm \citep{Isenberg1976} and depends on actin-binding proteins such as profilin \citep{Ozaki1984}. Shear flow also enhances depolymerization \citep{Kohama1988}. The polymerization and depolymerization of actin are also known to add an active stress contribution, included in active gel models \citep{Kruse2005, Joanny2013}. In \textit{Physarum}, these effects may contribute to initial channel formation \citep{Guy2011}. Describing a static actin network with the inner flow, with the Brinkman equation and including flow-induced depolymerization, a critical pressure below which channels appear has been predicted \citep{Guy2011}.

Besides, the actin mesh reorganizes through time: from a disorganized mesh in protoplasmic droplets \citep{Radszuweit2014, Brix1987_V}, the actin proteins form fibers, oriented longitudinally for new tubes and circumferentially for older tubes \citep{Stockem1994, Oettmeier2018}, enhancing the tension in the longitudinal direction \citep{Wohlfarth-Bottermann1976, Oster1984}. Such nematic ordering, however, breaks the isotropic symmetry assumed in poro-visco-elastic models in the constitutive laws for the gel. This alignment may be induced by the polymerization of actin or by applied stress \citep{Sherratt1993}.

Finally, the endoplasm is reported to have visco-elastic properties and a shear-dependent viscosity \citep{Sato1983}. With the varying elasticity of the ectoplasm, which is a compressible poroelastic material at short times, but viscous with lower effective oscillating Young's modulus \citep{Fessel2018}, it becomes evident that \textit{Physarum} mechanical properties remain to be characterized and modeled accordingly.

    \subsection{Calcium dynamics}
    The chemical properties of the calcium ion, $\rm{Ca}^{2+}$, make it an agent of choice for cells in sending spatial and temporal signals \citep{Carafoli2016}; calcium is involved in fertilization, development, apoptosis, or the control of cell contractility \citep{Berridge2000}. The first piece of evidence of the role of calcium in contraction was discovered in 1883: the interrupted cardiac rhythm of rat hearts resumed when exposed to calcium \citep{Carafoli2002}. However, its role differed between species: calcium activates contractions in animals \citep{Brookes2004} but inhibites them in plants \citep{Hepler2016}. In \textit{Physarum}, the effect of calcium on the actomyosin complex and contractions was debated for a long time \citep{Wohlfarth-Bottermann1979}, as of today, no clear consensus has been found \citep{Teplov2017}. Calcium ions can be found in several states within the cell: free in the cytoplasm, trapped in vacuoles \citep{Kato1977}, or bound to other proteins. Free calcium has a basal concentration of about 100nM \citep{Yoshiyama2009}, but as in other cells, much of the calcium is bound or trapped in vacuoles. 

    During a contraction, the calcium concentration also oscillates, a phenomenon observed as early as the 1970s with aequorin \citep{Ridgway1976a}. Furthermore, stress measurements in \textit{Physarum} tubes showed that calcium efflux varied with the same period but with an opposite phase: low stress at high calcium concentration \citep{Yoshimoto1981a}.

    Several preliminary studies first indicated calcium to be a promoter of contractile activity: extracts of actin and myosin were activated in the presence of low concentrations of calcium \citep{Kato1975}, an increase in the concentration of calcium near the contracting parts of the tubes \citep{Ridgway1976a}, and a large number of calcium-filled vacuoles near the relaxed parts or empty vacuoles near the contracting parts \citep{Kuroda1982}. These studies concluded that calcium ion acts in a similar manner in \textit{Physarum} and muscle cells \citep{Ridgway1976a}.

    However, several studies showed an opposite role of calcium in the actomyosin complex, coupled with ATP and membrane potential. First, improvement of myosin purification techniques showed that potential contaminations probably biased first results \citep{Kohama1980}. Furthermore, calcium concentration peaks precede the relaxation phase in microplasmodia \citep{Yoshiyama2009}. Finally, some myosins in \textit{Physarum} are only active when phosphorylated, but are inhibited by calcium \citep{Kawamichi2007, Kohama2016}. ATP, on the other hand, oscillates with the same frequency and in phase with tension \citep{Yoshimoto1981b}, which increases actomyosin tension to saturation, at around 1mM ATP \citep{Yoshimoto1984}. The presence of  a calcium-activated ATP-hydrolyzing enzyme might explain why ATP decreases when calcium increases \citep{Kamiya1981}. The measured voltage decrease is probably due to the presence of fragmin \citep{Sugino1983}.
    We propose the following simplified sequence of events to show the role of the different molecules on actomyosin contractions. Hyperpolarization of the membrane potential \citep{Hirose1980}, calcium level rises and ATP level falls \citep{Kamiya1981} leading to tension force decrease, by inhibiting myosins and lack of ATP, such that the cortex relaxes. Depolarization of the membrane potential \citep{Hirose1980}, calcium level decrease, and ATP production increase \citep{Yoshimoto1984, Hirose1980}, lead to myosin activation and contraction of the actomyosin cortex, generating tensile force.

    Overall, the role of calcium in \textit{Physarum} contraction  is still debated, but recent studies tend to show that it may rather inhibit contractions than promoting them, reviewed in \citep{Kohama2016}.
    
    \subsection{Fluid flows}
    Fluid mechanics play a crucial role in slime mold locomotion and nutrient transport, and flow equations can be easily derived in the plasmodial tubular organisation. For a single tube, the inner fluid is considered an incompressible Newtonian fluid. With fluid velocities ranging from $v=50 \mu\rm{m}/s$ to $v = 1 \rm{mm}/\rm{s}$, the tube's average radius $a_0=50 \mu \rm{m}$ and a kinematic viscosity of $\nu=\eta_s/\rho=6.4 \times 10^{-6} \rm{m}^2.\rm{s}^{-1}$, \citep{Swaminathan1997} with $\rho=1120\rm{kg}.\rm{m}^{-3}$ the cytoplasmic density \citep{Sato1983, Oettmeier2019a}, the Reynolds number is evaluated around $\mathrm{Re}\equiv v a_0 / \nu \simeq 10^{-3}$ \citep{Haupt2020, Oettmeier2019a, Alim2018}. In this limit, the Navier-Stokes equations reduce to the Stokes equation $\eta_s \nabla^2 \vec{v} = \vec{\nabla} p - \vec{f}$ with $f$ the external forces and $p$ the hydrostatic pressure. With no-slip boundary conditions, the flow profile is parabolic \citep{Haupt2020, Teplov1991, Kamiya.1950tvh, Bykov2009}, with small deviations in small tubes \citep{Romanovskii1995}. Thus, the average flow velocity is $v = -\frac{a_0^2}{4\eta_s}\partial_z p$, where the local pressure $p=\sigma_e + \sigma_a$ is generated by local contractions produced by visco-elastic stress $\sigma_e$ and active stress $\sigma_a$ \citep{Teplov1991}, and of the order of $1\rm{kPa}$ for a 1 cm long tube \citep{Teplov2017}. The tube being cylindrical, with total length $L$  much longer that the radius, $L \gg a_0$, the average flow rate $Q$ is typically expressed as the Hagen-Poiseuille equation  $Q = \frac{\pi a_0^4}{8\eta_s L}\Delta P = \Delta P/R$, analoguous to Ohm's law where $R$ is the hydraulic resistance and $\Delta P$ is the hydrostatic pressure difference between the two ends of the tube, generated by the local pressures. Finally, mass conservation for a tube segment is $\partial_t a^2 + \partial_z (a^2 v) = 0$.

\textit{Physarum} has a pulsatile flow, often referred to as ``shuttle stream" \citep{Kamiya.1950tvh, Stewart1959}, with a typical frequency of $\omega=0.05 \rm{Hz}$. The Womerseley number $\alpha$ is evaluated around $\alpha = a_0 \left(\frac{\omega}{\nu}\right)^{1/2} \simeq 10^{-3}$ \citep{Oettmeier2019a, Alim2018, Haupt2020}, indicating that viscous effects of the fluid dominate the pulsatile effects, and the flow profile is quasi-static with a parabolic shape.

Modifications of the above equations were proposed, first accounting for the peristaltic pumping by considering the tube shape deformation as a cosine function and explicitly calculating the contraction phase to shape relationship \citep{Lewis2017}, or extended to an entire network \citep{Alim2013}. Others explicitly coupled the active stress to the local concentration of a chemical agent and its advection through contractions \citep{Teplov1991, Alim2017, Julien2018, Kramar2021}, similar to poro-viscoelastic models. Finally, a feedback mechanism exists between the tube radius and the shear stress flow, leading to an adaptive tube radius, discussed in section 5.2.   

\section{MIGRATION}
In this section, we address the cell motility of \textit{Physarum}. First, we review the free migration of amoeboid-like cell or larger plasmodial network and then cover the response to external stimuli, such as chemicals, temperature, or light.

\begin{figure}
    \centering
    \includegraphics[scale=0.47]{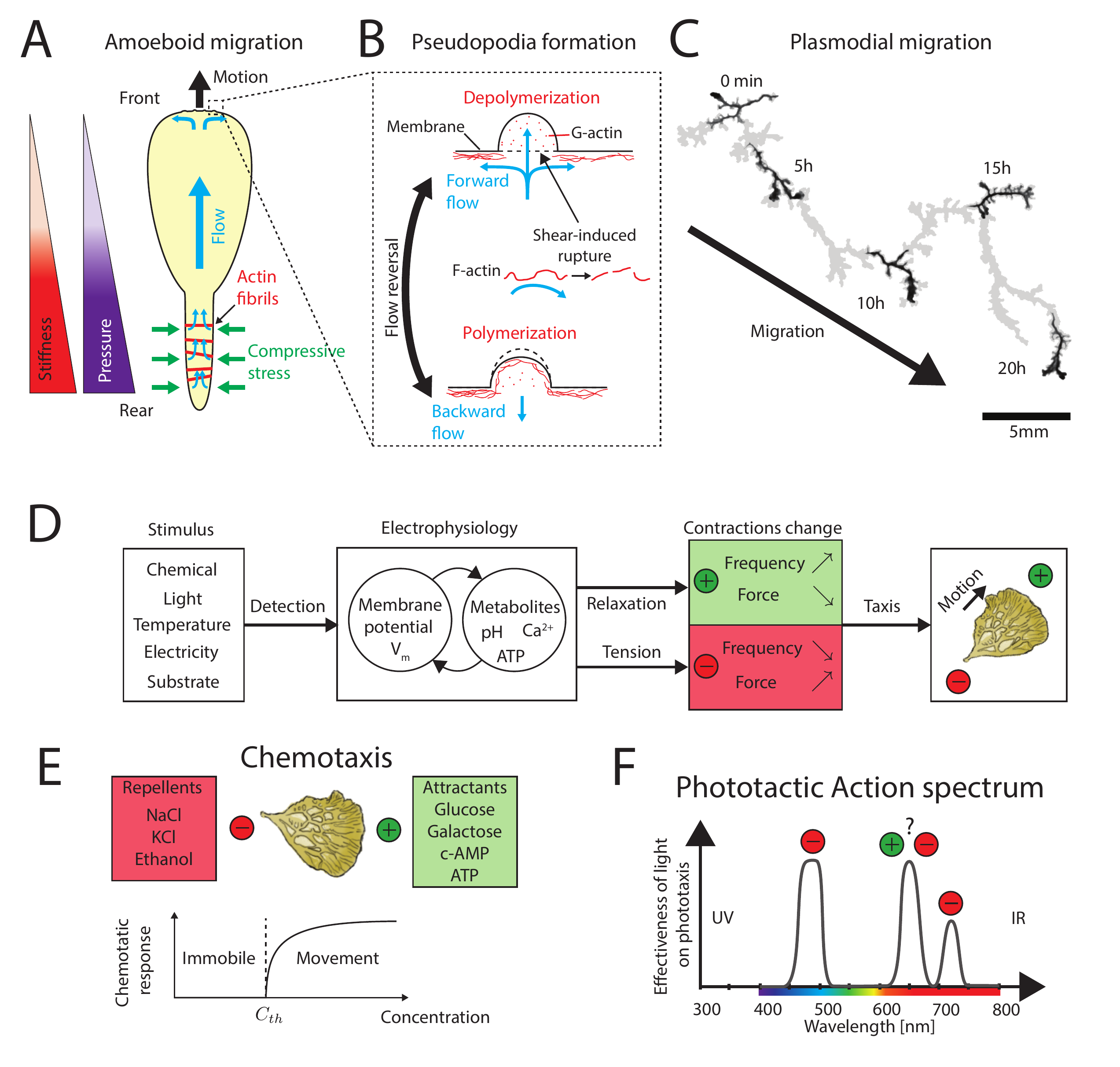}
    \caption{\textbf{Migration of \textit{Physarum}}. (A) Mechanism of amoeboid migration: stiffness gradient generates pressure gradient which triggers flow from rear to front. (B) At the front, pseudopodia form with shear-induced rupture of the F-actin cortex. Upon flow reversal, G-actin polymerizes and anchor the pseudopodia.  (C) At larger scales, the plasmodia moves randomly with several fronts, grey region indicates previously visited locations. (D) Proposed control pathway of taxis: a stimulus (light, chemical, etc.) is detected and alters the electrophysiology of the cell, which is translated through the contractions into migratory behavior. (E) Chemotaxis: above a threshold concentration $C_{th}$ of repellent or attractant, chemotactic response is observed. (F) Phototaxis in \textit{Physarum} can be represented with its action spectrum. Blue and far-red light are negative stimuli, while red light is reported as both a negative and positive stimulus, adapted from \citep{Hato1976}.}
    \label{fig:migration}
\end{figure}

    \subsection{Free migration}
    
\textit{Physarum} is a motile cell that performs amoeboid motion and reaches velocities of 2-3 cm/hour \citep{Kessler1982, Rieu2015}. However, its speed and motility depend on the size of the organism. Below 100 µm, a microplasmodium is immobile \citep{Rodiek2015b}, its rim oscillates under the effect of contraction waves in a spiral pattern, destroying the emergence of veins \citep{Takagi2008}. Around 100-200µm in size, radial symmetry breaks, and a front-rear axis is formed, with the rear distinguished by a local increase in the stiffness of the cytoskeleton. This symmetry-breaking appears to be controlled by the size and may originate from interfacial instability \citep{Zhang2019}.

With the establishment of an axis, the morphology of \textit{Physarum} evolves to resemble a tadpole \citep{Oettmeier2018, Rodiek2015b, Zhang2017}, see Fig.~\ref{fig:migration}A. Waves of peristaltic contraction appear: centripetal contractions start at the rear, generating inward stress perpendicular to the central axis \citep{Guy2011, Rieu2015}, and travel to the front \citep{Matsumoto2008, Ueda2011}. Birefringence microscopy also shows contractile actin fibrils, likely to cause this stress perpendicular to the central axis \citep{Rieu2015}. The contractions and the difference in cortex stiffness generate a pressure gradient, driving the liquid endoplasm from the rear to the front along the central axis \citep{Lewis2015, Ueda2011}. A large forward flow arrives at the front of the cell, likely causing actin fibers rupture due to high shear forces, observed for in-vitro reconstituted gels of skeletal muscle actin with \textit{Physarum} myosin \citep{Kohama1988, Maruyama1964}, Fig \ref{fig:migration}B. The cytoskeletal cortex breaks and pseudopodia are formed, and the cell membrane is pushed away \citep{Oettmeier2018}. Reversal of the flow stops the progression of the pseudopodia, and actin polymerizes again, gelifying the cytoskeleton, and forming new adhesion points \citep{Oettmeier2018}. Thus, the cell migrates in the direction of the new pseudopod.  The net movement of the cell is controlled by a flow asymmetry, larger from back to front than reversed \citep{Lewis2015}, a retractation time smaller than extension for the front \citep{Oettmeier2019a}, and by coordination of adhesion with contraction forces \citep{Rieu2015, Tanaka2012}. However, cell adhesion remains poorly characterized in \textit{Physarum}, with few studies showing the existence of filopodia-like membrane protrusions connected to the substrate by slime, forming focal adhesion points \citep{Brix1987_VII, Sesaki1997b}.

    Pseudopodia are reminiscent of cellular blebs \citep{Oettmeier2018, Oettmeier2019a}: a detachment of the plasma membrane following the rupture of the underlying cytoskeleton and then cytoplasmic flux leading to the polymerization of the actomyosin cortex \citep{Charras2008}. These blebs, also termed hyaline caps in other amobae \citep{Romanovskii1995}, have been observed in \textit{Dictyostelium discoideum} \citep{Yoshida2006} and are proposed as one of the mechanisms of cell migration. Blebs are inhibited by an increase in osmotic pressure that compensates for hydrostatic pressure in \textit{Dictyostelium} \citep{Yoshida2006} and in caffeine-induced blebs in \textit{Physarum} by external sucrose \citep{Kukulies1983}, indicating the crucial role of inner pressure in their formation.

    Several models tackled the locomotion of \textit{Physarum}, considering the mechano-chemical coupling of a chemical responsible for contraction (calcium, ATP, cAMP) and morphology \citep{Zhang2017, Kuroda2015}, mechanical models, including adhesion \citep{Tanaka2012}, hydrodynamics \citep{Romanovskii1995, Lewis2015, Oettmeier2019a, Lewis2017, Iima2012}, oscillators \citep{Iima2017}, or more phenomenological agent-based models \citep{Jones2010c}.

    Above a certain size, about 1~mm, several migration front appear \citep{Rodiek2015b}; see Fig.~\ref{fig:migration}C. These fronts evoke viscous fingering and emerge spontaneously below a critical curvature $\kappa = 2.5\, \rm{mm}^{-1}$, however, a model has yet to be proposed to describe this phenomenon at the mechanical level \citep{Baumgarten2015}. The fronts are initially large protoplasmic sheets with no apparent structure, but as the front advances, veins emerge and start to form a network \citep{Oettmeier2019a}. The veins in the direction of the shuttle stream are strengthened while others disappear. This reorganization, in the form of network coarsening has been modeled with an agent-based model \citep{Baumgarten2015} or by combining reinforcement models (see section 5) with reaction-diffusion and gel-sol models at the migration front \citep{Schenz2017}.

    The trajectory of an amoeba or a plasmodium is not uniform but akin to a run-and-tumble motion observed in other cells \citep{Rodiek2015a}. At short times, the center of mass follows a cycloidal path. At long time-scales, the trajectory alternates between phases of fast directed motion and phases of slow reorientations \citep{Rodiek2015a}, which coincides with a circular shape hence a loss of the previous direction \citep{Oettmeier2019a, Satoh1985}. The peculiarity of \textit{Physarum} lies on its self-avoiding walk due to the presence of slime, allowing it to dodge - if possible - regions already traversed \citep{Reid2013, Reid2012}. Migratory dynamics and morphology are correlated, as simple scalings where found: the velocity of the cell ($v$) increases almost linearly with its maximum thickness ($v \propto h_{\max}^{0.94}$), with allometric power-laws relating volume per unit length ($V$), length ($\ell \propto V^{0.66}$) and maximum thickness ($h_{\max} \propto V^{0.42}$) \citep{Kuroda2015}. 
    
    \subsection{Taxis}
    Like many cells, \textit{Physarum} can identify stimuli in its environment and orient itself accordingly, seeking food, avoiding poisons or light, to preserve itself and grow, a process referred to as \textbf{taxis}. Here, we will present the most common forms of taxis \textit{Physarum} is responsive to.
    
        \subsubsection{Chemotaxis}
       
        Whether it is to find food, to move away from harmful products, or to form colonies, many unicellulars have developed the ability to analyze their environment and its chemical composition \citep{Chet1976}. When a cell is able to orient itself and move according to chemical gradients, one speaks of \textbf{chemotaxis}, positive (resp.~negative) if the direction of movement follows (resp.~opposes) the positive concentration gradient. On the other hand, if the movement is not directed but velocity changes in this same chemical gradient, it is instead called \textbf{chemokinesis} \citep{Jakuszeit2021}.
        In \textit{Physarum}, first observations of chemotaxis date back to the work of Coman, who observed an attraction to glucose, but indifference to sucrose \citep{Coman1940, Carlile1970}. With the development of axenic cultures, many substances have been classified according to their attractiveness or repulsivity: sugars or carbohydrates \citep{Carlile1970, Ueda1975}, salts \citep{Durham1976, Terayama1977}, amino acids and nucleotides \citep{Hato1976b}, or phosphate compounds \citep{Ueda1975, Chet1977, Kincaid1978b, Knowles1978a, McClory1985}. In particular, glucose is reported as a chemoattractant, while NaCl as chemorepellent \citep{Patino-Ramirez2019}.

        Several methods have been used to measure the chemotactic response of \textit{Physarum}, mainly at the plasmodial stage \citep{Ueda1982}: at the macroscopic level with a double chamber measure pressure variations induced by plasmodial migration \citep{ Carlile1970, Terayama1977, Hato1976b, Chet1977, Kincaid1978b, Knowles1978a, McClory1985, Denbo1976, Kincaid1978a, Knowles1978b},  and at the mesoscopic level, with the measure of isometric tension in veins \citep{Ueda1976} or the electrophysiological response by measuring changes in the membrane electrical potential \citep{Ueda1975, Terayama1977, Ueda1985}. The chemotactic response of \textit{Physarum} depends on the concentration of the chemicals: above a certain threshold in concentration, a response is observed \citep{Ueda1975}, see Fig.~\ref{fig:migration}E. Conversely, high concentrations inhibit movement due to osmotic effects \citep{Knowles1978a, Denbo1976, Knowles1978b, Denbo1978}.

        How does \textit{Physarum} detect and then migrate in response to chemicals? We propose an electro-physiological feedback mechanism, inspired from \citep{Ueda1986}, see Fig.~\ref{fig:migration}D. First, external concentration changes at reception of a chemical substance at the cell membrane. Attractants are measured to hyper-polarize the membrane, while repellents depolarize the membrane potential $V_m$ \citep{Ueda1985}. Considering a repellent, the depolarization of the membrane increases pH and decreases free calcium \citep{Hirose1980} and subsequent contraction of actomyosin with ATP and calcium, see Fig.~\ref{fig:migration}D. Exposure to a chemo-attractant, such as glucose, can reverses this process to relax locally the tensions \citep{Natsume1992}. Still, the relationships between metabolites, membrane potential and stimuli needs to be entangled with further studies.

        Following substance detection, plasmodium tubes adjust their contraction oscillations: frequency increases with attractants, relaxing the tensile forces near the attractant, resulting in positive chemotaxis \citep{Kramar2021, Ueda1976}. Conversely, the frequency decreases with repellents, increasing the tensile forces and the cytoplasmic flow, pushing the cytoplasm towards the other portion and leading to negative chemotaxis \citep{Ueda1976, Chen2022_preprint, Miyake1994}.

        Finally, \textit{Physarum} migration is based on forming new tubes preferentially where the chemical gradient increases \citep{Reserva2021}. The morphology is also then influenced by the presence of chemicals: a compact morphology and slowed exploratory dynamics in high concentrations of nutrients (high nutrition) or salts (minimization of contact), in contrast to a trade-off between rapid migration, main branches, and optimal assimilation in nutrient-poor environments \citep{Patino-Ramirez2019, Nakagaki2000a, Takamatsu:2009kr, Latty.2011}.

        \subsubsection{Phototaxis}
        \textit{Physarum} thrives in humid and dark environments of forests. Light has been considered an aversive stimulus, which can affect its metabolism and behavior \citep{Sauer1982_book}.
        Two phenomena are distinguished: (i) \textbf{phototaxis}, the ability to evaluate the direction of light, and (ii) \textbf{photavoidance}, the ability to avoid light by moving \citep{Marwan2001}.
        The phototactic response can, thus, be measured through the contractile response of \textit{Physarum} tubes with isometric tension, or the overall mass displacement \citep{Hato1976, Kamiya1940}, resulting in an action spectrum mapping out the response of the organism for of several wavelengths, from UV to far-red regions see Fig.~\ref{fig:migration}F.
        Blue light (490 nm) is a robust negative stimulus \citep{Mori1986}. The organism finds the shortest path to minimize its exposure \citep{Nakagaki2007a} or moves away from the source by evacuating its mass \citep{Chen2022_preprint, Bauerle2020}. In particular, the frequency of contractions decreases \citep{Block1981, Baranowski1982, Takamatsu1997b}, as for a chemorepellent, while their amplitude increases \citep{Chen2022_preprint, Block1981}. With periodic exposure to blue light, one can synchronize the frequency of oscillations for a few cycles until the organism moves away (photoavoidance) \citep{Nakagaki1999a}.
        In contrast, green light (520nm) is reported to have no noticeable effect on phototaxis \citep{Rodiek2015a, Ueda1988}, which makes it useful for imaging.
        On the other side of the light spectrum, the effect of red light  remains more debated: \citep{Hato1976} reports a positive effect of red light (650nm) on phototaxis in Physarum, while other studies do not measure a significant effect \citep{Block1981, Schreckenbach1984}, or aversive stimulus for far-red (720nm) \citep{Hato1976}. However, red light can trigger the sporulation of a starved \textit{Physarum} \citep{Starostzik1995a, Starostzik1995b, Nakagaki1996a}. Notably, the synthesis of red (650 nm) and far-red (720 nm) light-detecting phytochromes are measured during starvation \citep{Lamparter2001, Ratzel2013}.

        \subsubsection{Galvanotaxis}
        
       Amoeba can be directed with an external electrical field, also termed \textbf{galvanotaxis}. Slime molds, in particular, were early reported to be affected by an external electrical field, with an electrical response upon mechanical or electrical stimuli \citep{Tasaki1950}, injury \citep{Kishimoto1958a} or pressure change \citep{Kishimoto1958b}, and migrating towards the cathode (negative pole) \citep{Anderson1951}. 
     Membrane potential is found to be between $-60$ to $-130 \mathrm{mV}$ with an electrical resistance around $0.1\,\mathrm{\Omega m^2}$ \citep{Hato1976b, Fingerle1982, Grigoriev2016}. The electric potential is proposed to be maintained by an extruding proton-pump \citep{Fingerle1982, Kuroda1981, Kuroda1989a} and to depend on the external pH \citep{Fingerle1982, Grigoriev2016}. Inhibition of the respiratory activity of \textit{Physarum} with KCN depolarizes the membrane \citep{Kuroda1989a}. The electrical potential was also measured to oscillate with the contractions of the cell wall, with a similar period \citep{Iwamura1949, Meyer1979, Zheng2015}. Inhibition of the contractions with neomycin also reduces electrical activity, which is proposed to inhibit the mechanosensitive calcium channels of the cell membrane, and showing a positive feedback mechanism between calcium entry and tube contractions, with membrane potential proportional to cell deformations \citep{Grigoriev2016}. An external electrical field has the advantage as a stimulus of being non-invasive, but its harmful effects on \textit{Physarum} remain to be quantified. Nevertheless, with such a tool, one has the potential to orient the migration of \textit{Physarum} with an attractive stimulus (cathode) without triggering developmental changes of the organism as with light or internal composition as with chemicals. More generally, several species of unicellulars, such as \textit{Amoeba proteus} \citep{Korohoda2000}, paramecium \citep{Ogawa2006}, \textit{Coleps hirtus} \citep{Daul2022}, or keratocytes \citep{Allen2013, Yang2013}, exhibit galvanotaxis, which can be used to study their associative memory \citep{DelaFuente2019}. 
     
        \subsubsection{Thermotaxis}
        \textit{Physarum} can orient itself in a temperature field, moving towards higher temperatures. As a plasmodial network, its minimum growth is around 15°C and can orient itself up to about 30°C, with optimal growth around 26°C \citep{Tso1975, Achenbach1980b, Wolf1997}, the temperature used for its culture. It can also detect variations of the order of 3°C \citep{Tso1975}. Thus, the temperature was often used as an external stimulus as it was much easier to control than chemicals and less harmful than light.
        Temperature variations affects the contraction cycles \citep{Nakagaki2000a, Achenbach1980b, Wolf1997, Wohlfarth-Bottermann1977a, Hejnowicz1980, Halvorsrud1995b, Takamatsu2004}: high temperature increases the contraction frequency, relaxes tensile forces \citep{Kamiya1959_Book}, resulting in positive taxis \citep{Achenbach1980b}. A power-law is proposed for the dependence of the contraction frequency $f$ on the temperature $T$, in the form $f(T) \propto (T-T_0)^b$, with $T_0= 273 \, \rm{K}$ and $b= 0.66$, \citep{Halvorsrud1995b}. There is also a trade-off between the antagonistic effects of light and temperature: intense illumination being a repellent, but locally increasing temperature being an attractant; however, illumination of \textit{Physarum} probably reduces other forms of taxa, resulting in negative taxis \citep{Nakagaki1996a}.
        
        Thermotaxis is not unique to \textit{Physarum} alone but is also observed in several unicellulars: paramecia were among the first thermotactic organisms reported \citep{Tawada1972, Nakaoka1977}, but also \textit{Dictyostelium discoideum} \citep{Poff1977}, sperm cells \citep{Bahat2003} or \textit{Escherichia Coli} \citep{Demir2012}.
        
        \subsubsection{Other forms of taxis}
        Applying a pressure on the cell wall or providing a stiffer or softer substrate often leads to a strong reaction of the tubular network of \textit{Physarum}, which often retracts from harmful situations. The stimulus being of mechanical origin, one speaks then of \textbf{mechanotaxis}.
        In particular, \textit{Physarum} can sense the local stiffness and relative deformation of the substrate on which it moves \citep{Umedachi2017, Murugan2021}, a form of \textbf{durotaxis} also present in animal cells and tissues \citep{Roca-Cusachs2013, Sunyer2020}. In the case of \textit{Physarum}, the inhibition of mechanosensory TRP channels cancels the mechanical detection \citep{Murugan2021}, a superfamily known to control mammalian cell mechanosensation, and to be a calcium and magnesium ion-channel \citep{Plant2014}.
        Several other stimuli have been applied to \textit{Physarum}: stretching of the tubes changes the phase of the contractions but not their frequency or amplitude \citep{Achenbach1982}, and the stretch also induce depolarization of the cell membrane \citep{Kamiya1981}. The response to gravity (\textbf{gravitaxis}), more precisely, the passage from 1G to 0G shows a transient response of the contractions then restoration to the initial oscillations \citep{Block1994}. Gravitaxis is especially useful for unicellulars to migrate vertically and find an ideal position in a water column without necessarily relying on phototaxis \citep{Hader1999}.

\section{COMPLEX NETWORK AND ORGANIZATION}
The interlaced, macroscopic network that \textit{Physarum} forms in its plasmodial stage undergoes continuous remodelling - coarsening as part of its migration and adapting in response to its environment. Here, geometric network characterization, network adaptation and the role of fluid flows for network remodelling and for transport is reviewed. 

\begin{figure}
    \centering
    \includegraphics[scale=0.3]{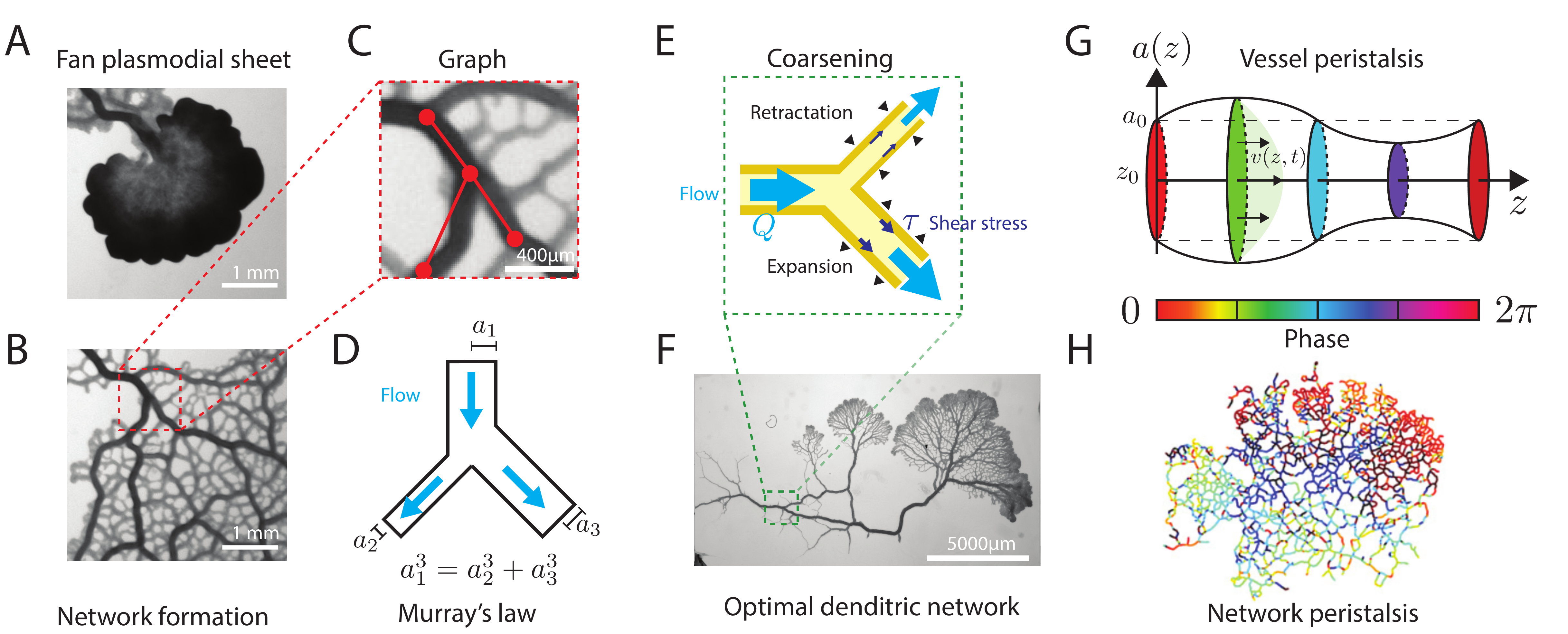}
    \caption{\textbf{\textit{Physarum} flow network}. (A) A plasmodial sheet, or fan, (B) reshapes into a network, (C) represented as a graph. (D) Minimizing viscous dissipation by flow at constant building cost predicts a geometric relation of tube radii, termed Murray's law. (E) A junction coarsens as tubes with small shear/flow retract while those with large shear/flow expand. (F) Optimal networks with dendritic shape are obtained through this coarsening. Phase gradient of contraction is observed along (G) a single tube and (H) across a large network, indicating peristaltic pumping, adapted with permission from \citep{Alim2013}.}
    \label{fig:network}
\end{figure}

    \subsection{Network organization}
\textit{Physarum}'s characteristic meshwork of tubes forms in the rear of an advancing migration front \citep{Oettmeier2018}, see Fig.~\ref{fig:network}A. At first, channels emerge in the large protoplasmic sheets of the migration front, see Fig.~\ref{fig:network}B, and as the front advances, tubes become clearly defined and the thin protoplasmic layer in between tubes thins away - a finely interlaced network forms, modeled as a graph, see Fig. \ref{fig:network}B, C. In network theory terms, the network is best described as a regular, planar graph of node degree three \citep{Baumgarten:2010fn}. Some nodes of degree four and dangling ends of degree one may be present \citep{Fessel:2012ch}, depending on the age and associated coarsening of the network after formation. Network tube radii have an average of $50$ µm \citep{Fricker.2017} and follow a log-normal distribution \citep{Baumgarten:2010fn}. As the network coarsens over time the width of the distribution shrinks \citep{Fricker.2017, Marbach:2016}. 

When \textit{Physarum}'s network is physically constrained in its growth and coerced to evacuating its material through a single tube, network geometry roughly follows Murray's law \citep{Akita.2016}. Murray's law follows from the assumption that vascular networks minimize viscous dissipation,  $2R Q^2$, governed by flow rate $Q$ and tube hydraulic resistance $R = \frac{8\mu L}{\pi a^4}$, at constant network building cost, typically proportional to the network volume \citep{Murray.19266e9}. Application of this optimization principle to a network junction predicts that the cubic radii of upstream daughter tubes are equal to the cubic radius of the downstream mother tube: $a_1^3 = a_2^3 + a_3^3$ \citep{Murray.19266e9}, see Fig. \ref{fig:network}D. Intriguingly the simple principle of minimal dissipation at constant network volume successfully describes vascular networks across the different forms of life, including not only \textit{Physarum} but also animals~\citep{West.1997,Kassab.1995} and plants~\citep{West.1997,McCulloh.2003} despite the very different biological makeup. This broad success of Murray's law suggests that vascular network adaptation in general is governed by the physics of laminar flow and \textit{Physarum} is likely a good model to understand for the physics of adaptive flow networks.

    \subsection{Adaptive network}
Murray's optimization principle conceptually describes the final or steady state architecture of a network predicting a tree-like \citep{Bohn.2007} or loopy morphology in the presence of fluctuations \citep{Durand.2006,Durand.2007,Corson.2010,Katifori.2010vpw,Ronellenfitsch.2019hi}. In this context, \textit{Physarum} networks constrained to connect localized food sources are found in their final state, when all food sources are still connected, to balance building cost in total network volume, average minimum distance between food sources and risk to disconnect network by failure of a single tube similar to man-build networks \citep{Tero:2010bx}. Such food-source connecting \textit{Physarum} networks are successfully predicted from initially inter-webbed networks, modelling food sources as randomly changing inflow-outflow pairs, with local tube diameter adaptation, $d\langle a\rangle/dt$, in response to the flow rate $Q$ pervading the tube \citep{Tero:2010bx, Tero.2006, Graewer.2015}. Here, computer scientists have found the exact functional dependence on the flow rate $Q$ to be minor for the final network pattern interconnecting food sources as long as it is a smooth monotonic function of itself \citep{Tero:2010bx, Tero.2006, Tero:2007hh, Tero:2008il, Gao2017}. That said the revelation that the flow driven adaptation laws provide a new algorithm to solve Steiner's problem received a lot of attention in computer science \citep{Gao2017, Gao2019}.

Without the constraint of food sources, \textit{Physarum} networks are even more dynamic. For example, a network will align its architecture in response to an imposed confinement within 3 hours \citep{Saiseau.2022}. The network coarsens and the overall transport within the remaining network increases by removing the small tubes serving redundant connections \citep{Baumgarten2015, Baumgarten:2010fn, Fricker.2017, Marbach:2016}. The coarsened networks, also termed dendritic, also exhibit lower energy consumption compared to interlaced meshlike networks as measured in oxygen consumption \citep{Takamatsu.2017}, see Fig.~\ref{fig:network}F. 
Laminar fluid flows drive tube adaptation and thus network reorganization as observed during coarsening, see Fig.~\ref{fig:network}E. Experiments quantifying the flow shear rate, $\tau=4 |Q|/\pi a^3$, revealed that tube radii follow the square of flow shear relative to a reference shear, $\tau_0$, albeit with a time delay,  \citep{Marbach.2022}. The time-delay was mathematically captured introducing a sensed shear $\tau_s$ governing time-averaged tube diameters $d\langle a\rangle/dt =\langle a\rangle/t_{\mathrm{adapt}}([\tau_s^2(\langle \tau \rangle)/\tau_0^2]-1)$ and following the time-averaged shear rate with a time delay $d\tau_s/dt=-1/t_{\mathrm{delay}}(\tau_s-\langle\tau\rangle)$. The square root dependence of tube adaptation dynamics on shear rate is predicted theoretically from local force balance on tube walls, described as active gels \citep{Marbach2023}. This mathematical framework now also revealed that individual tube flow shear is integrating the entire network architecture with the ratio of a tube resistance, $R$, relative to the remaining network resistance, $R_{\rm{net}}$, as the key parameter controlling if a tube is stable ($R<R_{\rm{net}}$) or coarsens ($R>R_{\rm{net}}$) \citep{Marbach.2022, Marbach2023}. This framework now explains observations of \textit{Physarum} discriminating parallel tubes by length ratio \citep{Mori2013} but also complex tube dynamics and cascading network reorganization as observed during network coarsening \citep{Marbach.2022}.
Thus, \textit{Physarum}'s cytoplasmic flows emerge to be at the heart of network adaptation.   
   
    \subsection{Peristaltic flows}
\textit{Physarum}'s plasmodial networks exhibits a back-and-forth flow of liquid endoplasm inside the tubes, referred to as a shuttle streaming \citep{Kamiya1981, Kamiya1940, Kamiya1959_Book}. Flow is generated by contractions of the actomyosin cortex lining the tube walls: by contracting tubes circumferentially, the cortex generates forces capable of displacing and, thus, moving fluid \citep{Rieu2015}.

In mathematical terms, the contraction driven flow within a single tube can be described by spatio-temporally varying tube radius $a(z,t)$ extending along $z$ to tube length $L$, filled with a fluid of velocity $v(z,t)$, see Fig.~\ref{fig:network}G. Tube radius dynamics can be modeled directly by balancing rhythmic contractions and visco-elastic restoring force \citep{Bauerle2020,Li.1993xte}, or by having the contraction emerge from a feedback between the visco-elastic stress and the active stress of the actomyosin cortex \citep{Teplov1991, Julien2018}. Contractions then generate a net flow of $\Delta Q = \pi \int_{z_0}^z dz' \frac{\partial a^2(z',t)} {\partial t} $. When spatially discretizing a tube as a combination of smaller successive tube segments, the phase of the contraction wave, $\varphi_j$, is to a good approximation constant in each tube segment and only varying smoothly across tube segments, such that the cross-sectional contraction can be modeled by $a_j^2(z,t) = a_{j,0}^2 + 2a_{j,0} \epsilon e^{i(\varphi_j - \omega t)}$. This formulation can be extended to all tubes of the network to represent a flow map of the organism, see Fig. \ref{fig:network}H. Due to mass conservation, not all phase patterns are allowed: the sum of contraction-driven flows must be zero: $\sum_{j \in \rm{tubes}}^N \Delta Q_j = 0$. This constrains the space of possible contraction states \citep{Fleig.2022} as one tube segment acting as a pump (dilating tube) must have a opposite source expelling fluid (contracting tube) \citep{Alim2013, Wilkinson2022}. 

In the context of network dominated morphologies, tubular contractions are the main driver of flows \citep{Fleig.2022}. Experimental observation reveal that both the entire tubular network and migration fronts contract rhythmically \citep{Yoshimoto1978b}. Cross-sectional contractions of the tube occur periodically, about every 1 to 2 minutes, and follow an imperfect phase gradient \citep{Hejnowicz1984}. Contractions are typically coordinated in a peristaltic pattern \citep{Alim2013, Iima2012} aligned along the longest axis of the organism or alternating between equally long directions if the network shape allows for more than one longest axis \citep{Alim2013}. Symmetric network shapes exhibit even richer contraction patterns \citep{Takagi2008, Fleig.2022}. The peristaltic wave's intrinsic wavelength matches network size \citep{Alim2013, Busson.2022}, therefore maximizing flow magnitude and net transport of matter. Standing-waves associated with slower migration velocity are observed in single-stranded plasmodia and in amoeba  \citep{Rodiek2015b}.

Cortex contractions self-organize and form patterns: in excised strands or networks, contractions resume within minutes and coordinate their phases \citep{Alim2013, Takagi2008, Yoshimoto1978a}. It was proposed that the contractions synchronization factor is transported with the endoplasmic flow \citep{Achenbach1980b}, although removing endoplasm from tubes was observed to not affect the persistence of contractile waves \citep{Samans1984}. A candidate for the synchronization factor is calcium  \citep{Teplov1991, Ridgway1976a, Julien2018, Wohlfarth-Boettermann1977b} also known to coordinate amoeba migration.

Analyzing the time-series of \textit{Physarum}'s cortex contractions the time-series decomposes into a fundamental frequency and a harmonic \citep{Proskurin2014, Avsievich2017}. This superposition of frequencies allows \textit{Physarum} to maximize the tube occlusion and, thus, flow rate \citep{Bauerle2020}. The fundamental frequency is also intrinsically modulated with a longer period of about 15-20 minutes \citep{Kuroda2015}. Notably, the fundamental frequency is affected by many chemicals or light stimuli, including respiration inhibitors \citep{Proskurin2014, Avsievich2017, Korohoda1983a}. A change in contractions is thereby linking cues in \textit{Physarum}'s environments to changes in network flows, thereby directly affecting the routing of transport within the network. 

    \subsection{Transport and signalling}
The coordinated contractions are generating long-ranged fluid flows \citep{Alim2013}, which are capable of transporting particles from a peripheral site across the entire network, within half a contraction period \citep{Nakagaki2000a} - before they are shuttled back in the second half of the contraction period.
Due to Taylor dispersion, flows may significantly increase particle diffusivity from their molecular diffusivity $\kappa$ to an effective diffusivity $\kappa+a^2v^2 / (48\kappa)$, larger than the net transport in peristaltic flow $v$ for a single closed tube \citep{Taylor.1953, Aris.1956}. 
Combined with the mixing of particle trajectories across a tube junction \citep{Haupt2020}, the effective dispersion unfolds, which can be augmented by network architecture coarsening \citep{Marbach:2016}. 

Flows transport both ions and larger molecules \citep{Zhang2017, Natsume.1993} but also organelles like nuclei \citep{Gerber2022}. Despite the mixing capabilities of the shuttle flow \citep{Haupt2020, Marbach:2016}, gradients in small molecules like ATP or calcium ions are reported \citep{Mori1986, Natsume.1993}. Spatial transcriptomics recently revealed regionality in the transcriptome associated with proliferation, syncytial substructures and localized environmental conditions \citep{Gerber2022}. These observations suggest that different parts of the networks may fulfill different tasks despite the lack of compartmentalization.

Spatial heterogeneity in \textit{Physarum} is also observed in the response to cues in its environment. When encountering a localized food source, contraction amplitude increases first close to the food source but then propagates with the speed of dispersive transport arising from Taylor dispersion within the network \citep{Alim2017}. Since the fluid flows are transporting the signal that increases contraction amplitude, the spread is heterogeneous, primarily following along large flow in big tubes. Note that the raise in contraction amplitude feeds back onto the flow, further boosting the flows and thus the transport of the signal \citep{Alim2017}. Likely the increase in amplitude stems from the signal triggering a softening of the tube walls' cortex, as the signal's routes grow quickly in tube diameter at the expense of drawing fluid volume from remaining tubes that in turn shrink in diameter \citep{Kramar2021}. Thus, flows transporting signals set up a food source specific pattern of thick and thin tubes forms that stores a memory of the food location and impacts the future behaviour of \textit{Physarum} \citep{Kramar2021}, one example of \textit{Physarum}'s memory capabilities allowing for complex behaviour.

\section{BEHAVIOR}

\textit{Physarum}'s characteristic plasmodial networks allow it to solve strikingly complex problems, like finding the shortest path through a maze \citep{Nakagaki:2000kp}, connecting food sources in efficient and robust transport networks \citep{Tero:2010bx}, balancing its optimal diet \citep{Dussutour:2010fu}, avoiding previously explored areas \citep{Reid2012} and arriving at informed decisions \citep{Latty.2011, Reid.2016}. The complexity of the problems solved seems at stark contrast with the simple network-like organization of the cell, and therefore intrigues to search for mechanistic insight of how behavioural complexity may emerge. Here, conceptualizing memory, information, decision making and learning may provide a framework for achieving insight in mechanisms that make life mount behaviours. 
 
    \subsection{Memory \& information}
The prerequisite to making informed decisions is the ability to store information from the past - to memorize. Broadly defined ``memory connotes the ability to encode, access, and erase signatures of the past in the state of a system'' \citep{Keim.2019}. In \textit{Physarum}, a whole set of different forms of memory have been discovered ranging from `external memory' \citep{Reid2012}, via shape memory in network morphology \citep{Kramar2021} to dynamical and chemical state memory \citep{Saigusa.2008, Vogel:2016bp, Boussard.2019} 

The term `external memory' was coined as \textit{Physarum} avoids its own trail of a thick mat of nonliving, translucent, extracellular slime, that it leaves behind when migrating \citep{Reid2012,Reid2013,Huynh2017}. \textit{Physarum} employs the extracellular slime's information of its past presence to solve navigation problems like the escape from a U-shaped trap \citep{Reid2012}. Therefore, the location of the extracellular slime literally encodes a trace of the past, that is read out as \textit{Physarum} can sense and avoid it. Here, the extended size of \textit{Physarum} exceeding the typical size of unicellulars is crucial, because the significant width of the slime trail ensures robust detection. 

Focusing on the organism itself, the network hierarchy of thick and thin tubes encodes the location of previously encountered food sources \citep{Kramar2021}. A memory that is read out as thick tubes redirect flow, transport and growth fronts \citep{Fleig.2022}. Memory is encoded into the network architecture as a soluble softening agent spreads from the location of food encounter with the fluid flows, softening and thus thickening tubes it reaches. Tubes not receiving softening agent shrink as fluid is drawn from them into the rapidly dilating tubes, within 5 to 15 minutes after food source encounter \citep{Kramar2021}. Given the analogous electrophysiological effects of a variety of environmental stimuli, it is likely that also other stimuli are embedded into the network architecture. Theoretical studies investigating memory formation within the context of adaptive networks minimizing dissipation cost at constant network volume found that memory formation intrinsically emerges from tubes shrinking and eventually pruning away and thus breaking ergodicity \citep{Bhattacharyya.2022}. Within this theoretical framework, it becomes also evident that the number of tubes at the stimulus encounter and further network coarsening after stimulus encoding are limiting the network architecture's memory \citep{Bhattacharyya.2023}.

The dynamical state of contractions has also been discussed as a mean to store the period of repeated environmental stimuli within dynamical systems synchronization of a set of to be identified  oscillators \citep{Saigusa.2008}. This notion is based on the training of a slow down in migration velocity upon threefold, periodic reduction in humidity and temperature at 30 to 90 minute intervals and recall of slow migration once the trained response had ceased \citep{Saigusa.2008}.

On much longer time scales, signatures of the past are stored in \textit{Physarum}'s chemical composition \citep{Vogel:2016bp, Boussard.2019}. When \textit{Physarum} specimen were trained to cross bridges of unfavourable salt conditions to reach food, they could transfer this behaviour to another specimen by fusion if it lasted not shorter than about 3 hours \citep{Vogel:2016bp}. The time constraint on fusion duration together with the observation of salt uptake \citep{Boussard.2019} points to a lasting change of chemical composition that is transferable by fluid flow and also outlasts one month dormant stages as sclerotia \citep{Boussard.2019}.

With the plethora of memory mechanisms in \textit{Physarum}, also information as the `measure of knowledge about the system' \citep{Shannon.1948} can be defined in a multitude of ways: as concentration of extracellular slime for `external memory', the amount or motion of bio-mass for network architecture memory, modulation of oscillations for dynamical state memory, or flow of concentration for chemical composition memory. 
Embracing quantifiable measures of information capacity and transfer with the framework provided by information theory, as already initiated in simple examples for \textit{Physarum} \citep{Ray.2019}, allows to assess not only how information is transferred and stored but also how a diverse set of information is processed to reach decisions and form behaviour.

    \subsection{Decision making \& learning}
    Decision-making is the ability to choose an action among different options, which in the context of a foraging slime mould as \textit{Physarum} could be binary yes or no decisions as to move or to stay put, to turn right or to turn left, to speed up or to speed down but also graded decisions as to speed up a bit, a lot or a whole lot. Contrary to `point-like' cellular organisms who can only be in one place at a time, \textit{Physarum}'s flexible morphology seems to allow it to circumvent decisions by being in multiple places at the same time. For example, \textit{Physarum} can cover different food patches at the same time and even distribute its mass to optimize its diet \citep{Dussutour:2010fu}. Yet, to arrive at this optimal distribution of body mass, decisions where to grow had to be performed. And indeed \textit{Physarum} reduces migration speed on rich substrate compared to poor substrate \citep{Latty.2009}, thereby linking chemotaxis to decisions. The rule to adapt migration speed to substrate quality also allows to explain \textit{Physarum}'s decision-making to exploit (an already acquired food source) or explore (for maybe much better food resources elsewhere) within a two-armed bandit setup \citep{Reid.2016}. In the two-armed bandit experiments, \textit{Physarum} was set to grow along two opposing arms each with a different distribution and amount of subsequent food sources. It was found to exploit, i.e.~not migrate along, an arm in proportion to its reward experienced through past sampling along that arm, also attracting further modeling work \cite{Kim2010}. Notably, in making decisions, stressors such as harmful light-exposure or deprivation of food, increase the speed to decide among food sources of different quality albeit at increasing the risk to choosing low quality food \citep{Latty.2011}.

In a similar manner, the complex problem to connect two food sources, as the shortest path through a maze \citep{Nakagaki:2000kp} or building robust transport networks \citep{Tero:2010bx}, can be reasoned to result from a simple mechanism: a food source triggers the release of a softening agent, preferentially maintaining and expanding nearby softened tubes at the expense of transport-wise less connected tubes, that are shrinking \citep{Kramar2021}. As the softening agent concentration is highest along the shortest route between food sources and further increases the flow along this connection by increasing contraction amplitude \citep{Kramar2021}, the selection of the shortest route to pervail results from increased flow and tube adaptation \citep{Tero:2010bx, Tero.2006, Marbach2023}. 

 While \textit{Physarum}'s repertoire of decisions is far from being fully documented, the question if \textit{Physarum} can adapt its decisions by past experiences, i.e.~learn, along the definition of reinforcement learning \citep{sutton1998reinforcement} is contemplated. In this context, \textit{Physarum} specimen were repeatedly distanced from a food source by a bridge of adverse chemical composition \citep{Boisseau:2016co}. With repeats the time to cross the unfavourable bridge decreased. The response to quickly cross the bridge could also be revoked to some extend after a period without adverse conditions leading the authors to the conclusion that \textit{Physarum} shows habituation \citep{Boisseau:2016co}, the simplest form of non-associative learning defined within psychobiology \citep{Rankin2009, Dussutour.2021fx}. Indeed, \textit{Physarum} seems to memorize past adverse conditions by its chemical composition \citep{Vogel:2016bp}. Now, the plethora for contrastive environmental cues, as outlined in section 4, opens up the possibility to also test for associative learning \citep{Shirakawa:2007ip}, although potential cross-effects of attractive and adverse stimuli need to be considered \citep{Murugan2021, Kunita.2017}. Here, quantitative studies of taxis in specific environments may be required to fully explore the breadth of potential learning capabilities in \textit{Physarum}.
    
Thus, \textit{Physarum} can make different decisions dictated by its architecture, chemical composition, or external stimuli. Yet, associating concepts traditionally developed for neuronal organisms with a single cell albeit its complex morphology remains challenging. How can we define a specific behavior in \textit{Physarum}? ``Behavior" denoting all the actions the organism performs to adapt to its environment or react to a stimulus. In this context, \textit{Physarum} seems able to alternate between several behaviors: a tug-of-war between exploration with a specific morphology \citep{Lee2018} or exploitation \citep{Latty.2009}. Such adaptability between different behaviors could naturally emerge from the different modes of contractions reported in \textit{Physarum} \citep{Fleig.2022}, as well as the oscillations in contraction \citep{Boussard2021}. Thus, can we say that \textit{Physarum} modifies its behavior according to the problems it faces and, therefore, acts in a ``smart" way? All our previous discussion seems to conclude that, even though it is not strictly a cognitive system, the slime mold still is able to efficiently react to adapt to stimuli. 
\endgroup

\section{CONCLUSION}

\textit{Physarum polycephalum} is an odd looking albeit fascinating organism to observe and study for the scientific and artistic communities, as well as the general public. Science popularization recently sparked the public's interest with large-scale experiments, popular science books, and online communication.

As physicists, it is a living system easy to manipulate, image, and quantify and an ideal organism to study organization across several spatial scales. Moreover, the multiple morphologies it adopts throughout its development or in response to its environment make it unique and amazingly versatile to tackle several fundamental questions, from cell movement to large-scale vasculature formation and optimization. Most fascinating are the emergent properties of \textit{Physarum} at the plasmodial stage, especially the features usually associated with multicellular organisms or multiple agent systems, despite it being a single-cell: hierarchical organization, information processing, memory, complex migratory behaviour, and decision-making capacity.

Research on \textit{Physarum} can be distinguished into its biological make-up, its mechanics, the resulting migration dynamics, network adaptation and smart behaviour. That said, the fascination may very well be embedded in such a distinction being futile, as all parts come into play and precisely that may make \textit{Physarum} such an enigmatic model system. 

\begin{summary}[SUMMARY POINTS]
\begin{enumerate}
\item Network tubes consist of inner fluid, rich in G-actin, and outer gel-like wall, dominated by filamentous F-actin driving rhythmic contractions of tube wall and ensuing flow of the inner fluid. Tubes reorganize as fluid and gel are converted into one another through actin polymerization and depolymerization. 
\item Light, chemical substances, electrical fields, temperature and substrate stiffness may act as attractant or repellent on \textit{Physarum}. Following stimulus detection the electrophysiology of the cell is altered changing contractions and thus cytoplasmic flows to redirect migration and alter network morphology.
\item Even in the absence of stimuli, networks reorganize in response to the flow shear exerted on the tube wall building efficient transport networks that transport and mix signals and resources. Here, flow shear rate integrates network architecture due to the global coupling imposed by conservation of fluid volume.
\item External slime trails, the pattern of network's tube diameter hierarchy and chemical composition store information of the organisms' past. Self-avoidance of slime trails, rerouting of mass flow by network hierarchy and mixing of chemical composition allows read out of stored information to make informed decisions for behaviour.
\end{enumerate}
\end{summary}

\begin{issues}[FUTURE ISSUES]
\begin{enumerate}
\item \textit{Physarum} exhibits spatially heterogeneous gene expression in functionally different body parts of the organism such as growing fronts, i.e.~fans, network tubes or retracting tubes. This contrasts with the strong fluid flows mixing cytoplasm content including organelles across the network. How may mechanochemical coupling of flows, transport and network morphology account for multi-functionality in a single-cell organism?
\item Stimuli in the environment directly alter the chemical and thus mechanical and dynamic state of the rhythmically contracting network. Can the networks' dynamic state represent individual behavioural states? How do dynamical states superimpose and determine the space of behaviours?
\item Network architecture is very dynamic and exhibits different properties over time. Is there a mapping of network architecture to network function and if so, when do networks change functional state?
\item \textit{Physarum} may store information in different forms. How long does information last and how are they processed individually and together to make decisions and mount behaviours?
\end{enumerate}
\end{issues}

\section*{DISCLOSURE STATEMENT}
The authors are not aware of any affiliations, memberships, funding, or financial holdings that might be perceived as affecting the objectivity of this review. 

\section*{ACKNOWLEDGMENTS}
We thank all members of the Alim lab for helpful discussions, and in particular S. Chen, F. Goirand, L. Schick, N. Schramma, L. Tr{\"o}ger for their feedbacks on the manuscript. We thank A. B{\"u}chl for Fig. \ref{fig:introduction}C, N. Schramma for Figure \ref{fig:biology}C, L. Tr{\"o}ger for Figure \ref{fig:migration}C, and L. Schick for Figure \ref{fig:network}F. This work was supported by the European Research Council (ERC) under the European Union's Horizon 2020 research and innovation program (grant agreement No.~947630, FlowMem) to K.A.

\bibliographystyle{ar-style4.bst} 
\bibliography{main}

\end{document}